# MANIFEST DUALITY IN LOW-ENERGY SUPERSTRINGS

W. SIEGEL*

*Institute for Theoretical Physics*
*State University of New York, Stony Brook, NY 11794-3840*

hep-th/9308133  27 Aug 1993

**ABSTRACT**

String theories inspire a new formalism for their low-energy limits. In this approach to these field theories, spacetime duality and stringy left/right handedness are manifest. Enlarged tangent-space symmetries allow the different fields (graviton, axion, Yang-Mills) to be treated as a single multiplet, even in the bosonic case, except for the dilaton (multiplet), which appears as the measure.

## 1. Introduction

*Duality*[1] is the global symmetry relating the fields $g_{mn}$ (metric), $b_{[mn]}$ (axion), and $A_{m\hat{a}}$ (abelian vectors) in string theory when these fields are constant in some directions. This symmetry is preserved in the low-energy limit. (In fact, it becomes larger: The elements of the group matrices generalize from integral to continuous.)

Our main result is that duality can be made manifest (linear) in a new formalism for (super) gravity + axion.[2] Fields are defined in a (2D+n)-dimensional space with manifest O(D,D+n) symmetry. This space is reduced to D dimensions by an O(D,D+n)-covariant constraint, whose solution "spontaneously" breaks this symmetry to O(n) (the global symmetry of the n abelian gauge vectors). However, for solutions that are independent of d of the dimensions (or by reducing the theory to D−d dimensions) the symmetry is partially restored, to O(d,d+n). The fields and their gauge symmetry are unified in the (2D+n)-dimensional space; the graviton and axion (and abelian gauge vectors) are parts of the same (2D+n)-dimensional field. (In the supersymmetric case there are also anticommuting coordinates, and thus a larger, graded, spontaneously broken symmetry.)

* Work supported by National Science Foundation grant PHY 9211367.
  Internet address: siegel@max.physics.sunysb.edu.

The major consequences are: (1) The field theory that represents the low-energy limit of string theory, when written in this formalism, more closely resembles string theory: The left/right handedness of string theory is built in, as the local tangent-space symmetry. A gauge can be chosen where the global part of this symmetry is preserved, so Feynman diagrams factorize into left- and right-handed parts. (2) Duality-related solutions to the field equations can be treated together. (3) The formalism may suggest new ways of understanding gravity (using larger tangent spaces), $\sigma$-models (with duality-covariant vertex operators), and string field theory (with an enlarged set of fields, corresponding to the new tangent-space gauge invariances).

The outline of this talk is: (1) We first consider the algebra of the string oscillators, which is manifestly duality covariant. As a result, duality is manifest in the hamiltonian (operator) formalism of string mechanics, but not in the lagrangian one. This affine Lie algebra defines the gauge group. (2) The (massless) fields are introduced as backgrounds for these operators. By restricting to the classical mechanics of strings, we obtain a field theory formalism which can be applied to the bosonic theory in all dimensions, and to the supersymmetric theories in (at least) D=3,4,6,10. (3) As usual, in the supersymmetric case some torsions can be defined without connections. They differ from the usual ones because the corresponding Lie algebras are affine, and they are all constrained to be trivial (vanishing or pure gauge). (4) Connections can be introduced to define covariant derivatives. (5) In constructing the complete set of torsions and curvatures, divergences play as important a role as curls. We explicitly construct curvatures and actions for the bosonic sector. The lagrangian for the metric, axion, and vectors is all contained within the (unique) curvature scalar, with the dilaton as the measure. (In the first-order formalism, the dilaton appears *only* as the measure.) (6) We consider the supersymmetric case in more detail for D=4. We find the (unconstrained) prepotential superfields, on which left/right handedness and duality are manifest. (In a final section added after the talk, we discuss modifications for nonabelian Yang-Mills.)

## 2. String oscillator algebra

We work in the hamiltonian (operator) formalism for the string because duality can be made manifest there. The oscillators of the string (including the Green-Schwarz formulation of the superstring) form an (abelian) affine Lie algebra:

$$[Z_M(1), Z_N(2)\} = i\delta'(2-1)\eta_{MN}$$

where "1" and "2" label the $\sigma$ coordinate of the string, $Z_M$ includes both the (super)spacetime momentum ($\sigma$-)density and the $\sigma$ derivative ($'$) of the spacetime coordinates, and $\eta_{MN}$ is the constant metric in the space of vectors that carry the superindices $M$:

$$Z_M = (P_{\mathbf{M}}, X'^{\mathbf{M}}, P_{\hat{m}} - \tfrac{1}{2}X'_{\hat{m}}) \quad \Rightarrow \quad \eta_{MN} = \begin{pmatrix} 0 & \delta_{\mathbf{M}}{}^{\mathbf{N}} & 0 \\ \delta^{\mathbf{M}}{}_{\mathbf{N}} & 0 & 0 \\ 0 & 0 & -\delta_{\hat{m}\hat{n}} \end{pmatrix}$$

(We have included the heterotic directions $\hat{m}$ by using only the right-handed oscillators.) Although $\eta_{MN}$ is constant, it is still generally covariant in the sense of ordinary spacetime, since it consists of Kronecker $\delta$'s. Duality is simply the global symmetry acting on the superindices $M$ that preserves the metric $\eta_{MN}$.

The local symmetry group elements $e^{-i\Lambda}$ are generated by the affine Lie algebra $Z_M$:
$$\Lambda = \int \lambda^M Z_M$$
(integrated over $\sigma$, where $\lambda$ is a function of $X^{\mathbf{M}}(\sigma)$). The gauge parameters include those for general (super)coordinate transformations ($\lambda^{\mathbf{M}}$), axion (2-form) gauge transformations ($\lambda_{\mathbf{M}}$), and abelian vector(-multiplet) gauge transformations ($\lambda^{\hat{m}}$). There is also the gauge invariance of the gauge invariance:
$$\delta \lambda^M = \partial^M \lambda \quad \Rightarrow \quad \delta \Lambda = 0$$

The affine Lie algebra defines a new Lie derivative:
$$[\Lambda_1, \Lambda_2] = i\Lambda_{[1,2]} \quad \Rightarrow \quad \lambda^M_{[1,2]} = \lambda^N_{[1} \partial_N \lambda^M_{2]} - \tfrac{1}{2} \lambda^N_{[1} \partial^M \lambda_{2]N}$$
(where brackets on indices indicate antisymmetrization).

We have used the fundamental identities (expressing the triviality of the winding modes):
$$A' = Z^M \partial_M A, \quad (\partial^M A)(\partial_M B) = \partial^M \partial_M A = 0$$
The Klein-Gordon-like equation $\partial^M \partial_M = 0$ (not the usual d'Alembertian because $\eta$ is off-diagonal) acts as a partner to the commutation of partial derivatives $[\partial_M, \partial_N\} = 0$. This condition is weaker than the vanishing of $\partial^{\mathbf{M}}, \partial_{\hat{m}}$ in
$$\partial_M = (\partial_{\mathbf{M}}, \partial^{\mathbf{M}}, \partial_{\hat{m}})$$
However, the most general solution to $(\partial^M A)(\partial_M B) = \partial^M \partial_M A = 0$ is of this form: The choice of a solution just chooses which derivatives, labeled $\partial^{\mathbf{M}}, \partial_{\hat{m}}$, are set to vanish. (I.e., the fields now depend on only $X^{\mathbf{M}}$.) Thus the solution of these conditions spontaneously breaks the manifest O(D,D+n) symmetry to O(n). However, if we choose fields such that an additional d of the D $\partial_M$ vanish, the symmetry is partially restored to O(d,d+n). (In the supersymmetric case the original symmetry is a graded one; e.g., for the heterotic string we have OSp(D,D+n|2D') for D' fermionic coordinates.)

## 3. Background fields

Background fields $e_A{}^M$ are introduced by covariantizing $Z_M$:

$$\Pi_A = e_A{}^M Z_M, \quad g_{AB} = e_A{}^M e_B{}^N \eta_{MN}$$

The algebra of $\Pi_A$ defines $\lambda^M$-covariant objects $F$:

$$[\Pi_A(1), \Pi_B(2)\} = i\delta'(2-1)\tfrac{1}{2}[g_{AB}(1) + g_{AB}(2)] + i\delta F_{AB}{}^C \Pi_C$$

$g$ and $F$ thus represent a background-dependent metric and structure functions for the new algebra. The first-class (Virasoro) constraints restrict the tangent-space metric and the gauge group to GL(D)⊗GL(D+n) on $A = (\mathcal{A}, \tilde{\mathcal{A}})$ (or in the supersymmetric case a graded symmetry, such as GL(D|D')⊗GL(D+n) for the heterotic string):

$$g_{\mathcal{A}\tilde{\mathcal{B}}} = 0$$

$$\Rightarrow \quad L_+ = \tfrac{1}{2} g^{\mathcal{A}\mathcal{B}} \Pi_{\mathcal{B}} \Pi_{\mathcal{A}}, \quad L_- = \tfrac{1}{2} g^{\tilde{\mathcal{A}}\tilde{\mathcal{B}}} \Pi_{\tilde{\mathcal{B}}} \Pi_{\tilde{\mathcal{A}}}; \quad L_+ + L_- = \tfrac{1}{2} Z^M Z_M$$

$L_+$ and $L_-$ are the generators of left- and right-handed conformal transformations, respectively, so $\Pi_{\mathcal{A}}$ is left-handed while $\Pi_{\tilde{\mathcal{A}}}$ is right-handed. (Others have considered SO(d)⊗SO(d+n)[3] and GL(d)[4] as gauge groups; here those groups can be obtained by partial gauge fixing. General linear groups have also been considered as Yang-Mills groups in ordinary gravity;[5] this is merely a reinterpretation of the early work by Cartan.[6]) Note that the generators of $\sigma$ reparametrizations ($\tfrac{1}{2} Z^M Z_M$) are background independent ($X'^{\mathbf{M}} P_{\mathbf{M}}$ is already generally covariant); the background couples only to $\tau$ reparametrizations.

The gauge transformations of the background fields are given by:

$$\delta \Pi_A = [-i\Lambda, \Pi_A] + \lambda_A{}^B \Pi_B, \quad \lambda_{\mathcal{A}}{}^{\tilde{\mathcal{B}}} = \lambda_{\tilde{\mathcal{A}}}{}^{\mathcal{B}} = 0$$

$$\Rightarrow \quad \delta e_A{}^M = (\lambda^N \partial_N e_A{}^M + e_{AN} \partial^{[M} \lambda^{N]}) + \lambda_A{}^B e_B{}^M$$

$$\delta g_{AB} = \lambda^M \partial_M g_{AB} + \lambda_{(AB]}$$

The vielbein $e_A{}^M$ thus transforms as a cross between (but not a product of) a covariant and a contravariant world-vector. This is to preserve the super spacetime metric $\eta$, which is constant. On the other hand, the tangent-space metric $g$ is not constant (but can be gauged to constant by $\lambda_{(AB]}$). This is the opposite of the situation in the usual vierbein formulation of gravity. (The usual spacetime metric can be found in $L_\pm$ as part of the coefficients of $Z_M Z_N$.)

In the conventional gauge for the tangent-space group $\lambda_A{}^B$, $e_A{}^M$ reduces to the usual fields $e_{\mathbf{a}}{}^{\mathbf{M}}$ (graviton), $b_{\mathbf{MN}}$ (axion), and $A_{\mathbf{M}\tilde{a}}$ (vectors).

The dilaton appears as the integration measure, which can't be constructed covariantly from $e_A{}^M$ ($det\ e_A{}^M$ is not a density):

$$\delta \Phi^2 = \partial_M (\lambda^M \Phi^2)$$

# 4. Torsions and constraints

We now specialize to the (heterotic) superstring, in the Green-Schwarz formalism. (Background fields for the Green-Schwarz formalism were first considered by Witten,[7] and in the string hamiltonian formalism by Shapiro and Taylor.[8]) The indices take the values

$$M = (\mathbf{M}, \tilde{\mathbf{M}}, \hat{m}) = (\mu, m; \tilde{\mu}, \tilde{m}; \hat{m}), \quad A = (\mathcal{A}, \tilde{\mathcal{A}}) = (\alpha, a, \tilde{\alpha}; \tilde{a}, \hat{a})$$

$X^{\mathbf{M}}$ now has fermionic parts $\Theta^\mu$ as well as the bosonic parts $X^m$. (Similarly, the new tangent-space indices $\alpha, \tilde{\alpha}$ are fermionic.) We choose $\Pi_\alpha$ as the second-class constraints. This leads to further restrictions on the tangent-space metric and gauge group:

$$g_{\alpha\beta} = 0, \quad \lambda_\alpha{}^b = \lambda_\alpha{}^{\tilde{\beta}} = \Pi_\alpha{}^\beta{}_\delta{}^\gamma \lambda_\gamma{}^\delta = 0$$

where $\Pi_\alpha{}^\beta{}_\delta{}^\gamma$ is a projection operator constraining $\lambda_\alpha{}^\beta$ to be Lorentz⊕scale⊕internal.

The connection-independent torsions are:

$$F_{\alpha\beta}{}^{\tilde{\mathcal{C}}}, \ F_{\alpha\beta}{}^{c,\tilde{\gamma}}, \ \Pi_{\alpha\beta}{}^\gamma{}_\delta{}^{\epsilon\zeta} F_{\epsilon\zeta}{}^\delta, \ F_{\alpha\tilde{\mathcal{B}}}{}^{c,\tilde{\gamma}}, \ \Pi_\alpha{}^\gamma{}_\delta{}^\epsilon F_{\epsilon\tilde{\mathcal{B}}}{}^\delta$$

($\Pi_{\alpha\beta}{}^\gamma{}_\delta{}^{\epsilon\zeta}$ is another projection operator serving the same purpose. "$c, \tilde{\gamma}$" means $c$ or $\tilde{\gamma}$.) By requiring that the constraint algebra takes the usual form,

$$[L_-, \Pi_\alpha] \approx 0, \quad \{\Pi_\alpha, \Pi_\beta\} \approx i\delta F_{\alpha\beta}{}^c \widehat{\Pi}_c$$

(where "≈" means modulo the constraints) for some $\widehat{\Pi}$, we find the constraints on the background fields:

$$F_{\alpha\beta}{}^{\tilde{\mathcal{C}}} = F_{\alpha\beta\gamma} = F_{\alpha\tilde{\mathcal{B}}}{}^{c,\tilde{\gamma}} = 0, \quad F_{\alpha\beta}{}^{c,\tilde{\gamma}} = \gamma^{\check{d}}_{\alpha\beta} F_{\check{d}}{}^{c,\tilde{\gamma}}$$

(for some $F_{\check{d}}{}^{c,\tilde{\gamma}}$), where $\gamma^{\check{d}}_{\alpha\beta}$ are the usual $\gamma$ (or $\sigma$) matrices. The Bianchi identities then constrain the remaining connection-independent torsions:

$$\Pi_{\alpha\beta}{}^\gamma{}_\delta{}^{\epsilon\zeta} F_{\epsilon\zeta}{}^\delta = \Pi_\alpha{}^\gamma{}_\delta{}^\epsilon F_{\epsilon\tilde{\mathcal{B}}}{}^\delta = 0$$

In the conventional gauge these constraints give the usual result:[9]

$$H_{\alpha\beta\gamma} = 0; \quad \mathcal{F}_{\alpha\beta\hat{c}} = 0, \quad c_{\alpha\beta}{}^c = -H_{\alpha\beta}{}^c = \gamma^c_{\alpha\beta};$$

$$H_{\alpha bc} = c_{\alpha(bc)} = 0, \quad \mathcal{F}_{\alpha b}{}^{\hat{c}} = \gamma_{b\alpha\beta} W^{\beta\hat{c}}$$

where $c$, $H$, and $\mathcal{F}$ are the usual duality-noncovariant superfield strengths for the supergraviton (connection-free torsion), superaxion, and superMaxwell fields. ($W$, which originally appeared as the gauge field $e_{\tilde{\alpha}}{}^{\hat{m}}$, now becomes the usual spinor superfield strength.)

# 5. Covariant derivatives

We first define the covariant derivative in the naive way:

$$\nabla_A = e_A + \omega_{A\mathcal{B}}{}^{\mathcal{C}} G_{\mathcal{C}}{}^{\mathcal{B}} + \omega_{A\tilde{\mathcal{B}}}{}^{\tilde{\mathcal{C}}} G_{\tilde{\mathcal{C}}}{}^{\tilde{\mathcal{B}}}, \quad e_A = e_A{}^M \partial_M$$

$G_{\mathcal{C}}{}^{\mathcal{B}}$ and $G_{\tilde{\mathcal{C}}}{}^{\tilde{\mathcal{B}}}$ are the generators of the left- and right-handed GL symmetries, which act on all tangent-space indices. However, there is a new definintion of the torsion, determined by the new Lie derivative:

$$\lambda^A_{[1,2]} = \lambda^B_{[1} \nabla_B \lambda^A_{2]} - \tfrac{1}{2} \lambda_{[1B} \nabla^A \lambda^B_{2]} + \lambda^B_1 \lambda^C_2 T_{BC}{}^A$$

$$\Rightarrow \quad T_{AB}{}^C = F_{AB}{}^C + (\omega_{[AB)}{}^C + \tfrac{1}{2} \omega^C{}_{[AB)})$$

Consequently, the connection constraints don't fix all the connections:

$$\omega\text{-dependent } T_{AB}{}^C = \nabla_A g_{BC} = 0$$

$$\Rightarrow \quad \omega_{A\mathcal{B}}{}^{\tilde{\mathcal{C}}} = -F_{A\mathcal{B}}{}^{\tilde{\mathcal{C}}}, \quad \omega_{A(\mathcal{BC})} = -e_A g_{\mathcal{BC}}, \quad \omega_{[\mathcal{ABC})} = -\tfrac{1}{3} F_{[\mathcal{ABC})}$$

(In the bosonic case, all the torsions are $\omega$-dependent; in the supersymmetric case, the other torsions are fixed for other reasons, as described in the previous section.) Integration by parts for the measure $\Phi^2$ defines another new torsion and connection constraint:

$$\widetilde{T}_A \equiv \Phi^2 \overleftarrow{\nabla}_A \Phi^{-2} = 0$$

$$\Rightarrow \quad \omega_{BA}{}^B = -\tilde{F}_A \equiv -\Phi^2 \overleftarrow{e}_A \Phi^{-2}$$

(The backward arrow means the partial derivatives and GL generators act backwards, even on the fields contained in $\nabla$.) As the analog of the usual relation

$$[\nabla_A, \nabla_B\} A = T_{AB}{}^C \nabla_C A$$

from $[\partial_M, \partial_N\} = 0$, we have from $\partial_M \partial^M = 0$

$$\nabla_A \nabla^A A = -\widetilde{T}_A \nabla^A A$$

Since not all connections are defined, only certain kinds of differentiation are covariant: These include the new Lie derivative, divergences of left- and right-handed vectors, gradients of scalars, and gradients of opposite-handed vectors (e.g. $\nabla_A V_{\tilde{\mathcal{B}}}$), as well as combinations of these, and certain spinor derivatives in the supersymmetric case.

# 6. Curvatures and actions (bosonic)

Because of the condition $\partial^M \partial_M = 0$, traces are now a fundamental part of the definition of curvatures as well as torsions:

$$R_{A\tilde{B}} = 2R_{\tilde{C}A\tilde{B}}{}^{\tilde{C}} = 2R_{C\tilde{B}A}{}^{C}$$
$$= (e_{\tilde{B}}\widetilde{F}_A - F_{\tilde{B}A}{}^C \widetilde{F}_C) - (e_C F_{\tilde{B}A}{}^C - F_{C\tilde{B}}{}^{\tilde{\mathcal{E}}} F_{\tilde{\mathcal{E}}A}{}^C)$$
$$R = -\tfrac{1}{2} R_{AB}{}^{AB} = \tfrac{1}{2} R_{\tilde{A}\tilde{B}}{}^{\tilde{A}\tilde{B}}$$
$$= e_A \widetilde{F}^A + \tfrac{1}{2}\widetilde{F}_A{}^2 + \tfrac{1}{2} e_A e_B g^{AB} - \tfrac{1}{2} F_{AB\tilde{C}}{}^2 - \tfrac{1}{432} F_{[ABC]}{}^2 + \tfrac{1}{8}(e^A g^{BC})(e_B g_{AC})$$

These satisfy the Bianchi identities:

$$\nabla^{\tilde{B}} R_{A\tilde{B}} + \nabla_A R = \nabla^B R_{B\tilde{A}} - \nabla_{\tilde{A}} R = 0$$

The bosonic action is then simply:

$$S = -\int d^D x\ \Phi^2 R$$

This action implies the field equations:

$$-\delta S = \int R\ \delta\Phi^2 + \Phi^2 R^{A\tilde{B}} e_{\tilde{B}}{}^M \delta e_{AM} \quad\Rightarrow\quad R = R_{A\tilde{B}} = 0$$

The action can also be expressed in 1st-order form:

$$S = \int d^D x\ \Phi^2 \tfrac{1}{4}(R_{AB}{}^{AB} - R_{\tilde{A}\tilde{B}}{}^{\tilde{A}\tilde{B}})$$

$$-R_{AB}{}^{AB} = \tfrac{1}{2}(\tfrac{1}{2}\omega_B{}^{[AB)2} - 2e_A\omega_B{}^{[AB)} - \omega_B{}^{[AB)} e^C g_{AC})$$
$$- \tfrac{1}{2}(\tfrac{1}{2}\omega_B{}^{(AB]2} + \omega_B{}^{(AB]} e^C g_{AC}) + \tfrac{1}{12}(\tfrac{1}{2}\omega^{[ABC)2} + \tfrac{1}{3}\omega^{[ABC)} F_{[ABC)})$$
$$+ \tfrac{1}{12}(\tfrac{1}{2}\omega^{(ABC]2} + \tfrac{1}{2}\omega^{(ABC]} e_{(A} g_{BC]}) + (\tfrac{1}{2}\omega^{\tilde{A}BC2} + \omega^{\tilde{A}BC} F_{\tilde{A}BC})$$

Note that in the 1st-order form the dilaton appears only as the measure. (We can also add a cosmological term $\int d^D x\ \Phi^2$.)

For purposes of perturbation theory we can write

$$e_A{}^M \equiv \langle e_A{}^M \rangle + h_A{}^B \langle e_B{}^M \rangle, \quad \Phi^2 \equiv 1 + \phi$$

Using the tangent-space gauge invariance and the constraint $g_{A\tilde{B}} = 0$, we can choose $h_{A\tilde{B}}$ as the only independent field (its indices raised and lowered with the flat metric $\eta_{AB} \equiv \langle g_{AB}\rangle$). Using the gauge-fixing functions

$$f_A \equiv d^{\tilde{B}} h_{A\tilde{B}} + \tfrac{1}{2} d_A \phi, \quad f_{\tilde{A}} \equiv -d^A h_{A\tilde{B}} + \tfrac{1}{2} d_{\tilde{B}} \phi \qquad (d_A \equiv \langle e_A{}^M\rangle \partial_M)$$

the kinetic terms from $L - \tfrac{1}{2} f_A{}^2 + \tfrac{1}{2} f_{\tilde{A}}{}^2$ are simply

$$-\tfrac{1}{4}\phi\Box\phi - \tfrac{1}{2} h^{A\tilde{B}}\Box h_{A\tilde{B}}, \qquad \Box = d^A d_A = -d^{\tilde{A}} d_{\tilde{A}}$$

so the propagators are string-like. We can also write the rest of the action in terms of just $h_{A\tilde{B}}$, $\phi$, and $d_A$. We then see that the Feynman diagrams involve only $d_A$'s and $\eta_{AB}$'s (i.e., $\eta_{AB}$'s and $\eta_{\tilde{A}\tilde{B}}$'s): no objects which mix left- and right-handed indices. As a result, the (integrand of) each Feynman diagram factorizes into left- and right-handed parts (even though $d_A$ and $d_{\tilde{A}}$ aren't independent).

# 7. D=4

In D=3 the constraints determine the vielbein in terms of just the spinor part, as usual,[10] $e_A{}^M \to e_\alpha{}^M$, which is now extended to include the axion (multiplet) by virtue of its extended index $M$. In D=4 the constraints further determine the vielbein in terms of a prepotential:

$$\Pi_\alpha = A_\alpha{}^\mu e^W Z_\mu e^{-W}, \quad \Pi_{\dot\alpha} = A_{\dot\alpha}{}^{\dot\mu} e^{-\overline{W}} Z_{\dot\mu} e^{\overline{W}}, \quad W = \int W^M Z_M$$

$$e_\alpha = A_\alpha{}^\mu e^w \partial_\mu e^{-w}, \quad w = W^M i\partial_M$$

Again this result is similar to pure supergravity in ordinary superspace,[11] except that the index on the prepotential $W^M$ is extended to include the additional fields, and the expansion of the exponentials as multiple commutators contains extra terms due to the use of the new Lie bracket. (A similar result has been obtained by Berkovits,[12] but duality was not manifest[13] because the solution to the constraints was expressed differently.[14])

The generalization of the gauge group is also analogous to the pure supergravity case, with modifications from the extended indices and new Lie bracket:

$$e^{W'} = e^{-i\Lambda} e^W e^{i\Xi}, \quad \Xi = \int \xi^M Z_M, \quad \partial_\mu \xi^M = 0 \; except \; \partial_\mu \xi^\nu \neq 0$$

We thus have the analog of the gauge invariance of the gauge invariance:

$$W'^M = W^M + \partial^M \zeta$$

As in pure supergravity, these new transformations replace the old when using the $\Lambda$-invariant combination of the fields:

$$e^U \equiv e^{\overline{W}} e^W$$

The original gauge invariances can be used to eliminate all fields except the prepotential $U^M$, which transforms under only the new invariances $\Xi$ and $\zeta$. (As for the bosonic case, the dilaton multiplet appears in a separate superfield.)

The (superfield) curvatures are now $R$ (chiral) and $R_{\tilde{\mathcal{A}}}$ (real). The component Ricci scalar $R$ appears at order $\theta^2$ in the superfield $R$ (as in pure supergravity), while the component $R_{a\tilde{\mathcal{B}}}$ (which contains the remaining field equations) appears at order $\theta\bar\theta$ in the superfield $R_{\tilde{\mathcal{A}}}$. The latter superfield exhibits the direct-product (left$\otimes$right) structure of the states: $R_{\tilde{\mathcal{A}}}(\theta,\bar\theta) \leftarrow V(\theta,\bar\theta) \otimes A_{\tilde{\mathcal{A}}}$ (vector multiplet$\otimes$(vector$\oplus$scalars)).

## 8. Epilog

The true low-energy limit of the heterotic superstring (at least for interesting compactifications of the 16 extra dimensions) includes nonabelian Yang-Mills. After this talk was given we generalized the above methods to this case. It requires only the corresponding generalization of the affine Lie algebra:

$$[Z_M(1), Z_N(2)\} = i\delta'(2-1)\eta_{MN} + i\delta(1-2)f_{MN}{}^P Z_P$$

The group structure constants $f$ satisfy the usual Jacobi identities. If we define the unbroken duality group still to be the invariance O(D,D+n) (with n now the number of generators of the nonabelian Yang-Mills group) of the metric $\eta$, then $f$ transforms covariantly: After an O(D,D+n) transformation, it still has the algebraic properties of the structure constants of the Yang-Mills group (crossed with 2D U(1)'s). The solution of $\partial^M \partial_M = 0$ that spontaneously breaks this symmetry can be chosen such that the only nonvanishing components of $f$ are $f_{\hat{m}\hat{n}}{}^{\hat{p}}$. Then, when choosing the fields to be independent of d coordinates, the restored O(d,d+n) symmetry includes the nonabelian generators.

We then find the corresponding modification in the definition of the new Lie bracket:

$$\lambda_{[1,2]}^M = \lambda_{[1}^N \partial_N \lambda_{2]}^M - \tfrac{1}{2}\lambda_{[1}^N \partial^M \lambda_{2]N} + \tfrac{1}{2}\lambda_1^N \lambda_2^P f_{NP}{}^M$$

$F$ (and thus the torsions) are modified simply by $F \to F + f$, and the vielbein transformation law by $\delta e_A{}^M \to \delta e_A{}^M + e_A{}^P \lambda^N f_{NP}{}^M$. The low-energy action then has the same form as before, the nonabelian terms being automatically included in the redefinition of $F$. All these changes follow directly from the change in $[Z, Z\}$. Effectively, we have just changed our choice of vacuum, writing $\Pi_A = e_A{}^M \langle \Pi_M \rangle$, where $\langle \Pi \rangle$ is our new $Z$, which can be expressed in terms of some $\langle e \rangle$ and abelian $Z$. However, working directly in terms of the nonabelian $Z$ simplifies algebra, and allows the use of arbitrary representations (e.g., bosonic or fermionic) of this affine Lie algebra. If Lorentz generators are introduced as additional generators in the affine Lie algebra instead of as an ordinary[15] (but local in $\sigma$) Lie algebra (i.e., they then have a $\delta'$ term in their algebra), the low-energy action will then include Lorentz Chern-Simons terms.

These results can also be applied to the infinite-dimensional (spontaneously broken) Yang-Mills groups (n=$\infty$) which can occur when considering directly all the possible low-energy limits of a specific compactified string.[16]

## Acknowledgments

I thank Amit Giveon and Martin Roček for discussions related to the topic of the last section.

# REFERENCES

[1] K. Kikkawa and M. Yamasaki, *Phys. Lett.* **149B** (1984) 357;
N. Sakai and I. Senda, *Prog. Theor. Phys.* **75** (1986) 692;
V.P. Nair, A. Shapere, A. Strominger, and F. Wilczek, *Nucl. Phys.* **B287** (1987) 402;
B. Sathiapalan, *Phys. Rev. Lett.* **58** (1987) 1597;
R. Dijkgraaf, E. Verlinde, and H. Verlinde, *Comm. Math. Phys.* **115** (1988) 649;
K.S. Narain, M.H. Sarmadi, and E. Witten, *Nucl. Phys.* **B279** (1987) 369;
P. Ginsparg, *Phys. Rev.* **D35** (1987) 648;
P. Ginsparg and C. Vafa, *Nucl. Phys.* **B289** (1987) 414;
S. Cecotti, S. Ferrara, and L. Girardello, *Nucl. Phys.* **B308** (1988) 436;
R. Brandenberger and C. Vafa, *Nucl. Phys.* **B316** (1988) 391;
A. Giveon, E. Rabinovici, and G. Veneziano, *Nucl. Phys.* **B322** (1989) 167;
A. Shapere and F. Wilczek, *Nucl. Phys.* **B320** (1989) 669;
M. Dine, P. Huet, and N. Seiberg, *Nucl. Phys.* **B322** (1989) 301;
J. Molera and B. Ovrut, *Phys. Rev.* **D40** (1989) 1146;
K.A. Meissner and G. Veneziano, *Phys. Lett.* **267B** (1991) 33;
A.A. Tseytlin and C. Vafa, *Nucl. Phys.* **B372** (1992) 443;
M. Roček and E. Verlinde, *Nucl. Phys.* **B373** (1992) 630;
J.H. Horne, G.T. Horowitz, and A.R. Steif, *Phys. Rev. Lett.* **68** (1992) 568;
A. Sen, *Phys. Lett.* **271B** (1992) 295;
A. Giveon and M. Roček, *Nucl. Phys.* **B380** (1992) 128.
[2] W. Siegel, *Phys. Rev.* **D47** (1993) 5453; Superspace duality in low-energy superstrings, Stony Brook preprint ITP-SB-93-28 (May 1993), to appear in *Phys. Rev.* **D**.
[3] M.J. Duff, *Nucl. Phys.* **B335** (1990) 610.
[4] J. Maharana and J.H. Schwarz, *Nucl. Phys.* **B390** (1993) 3.
[5] E.A. Lord, *Phys. Lett.* **65A** (1978) 1;
A. Trautman, *Czech. J. Phys.* **29** (1979) 107;
A. Komar, *Phys. Rev.* **D30** (1984) 305;
R. Percacci, Role of soldering in gravity theory, *in* Differential geometric methods in theoretical physics, Aug. 20-24, 1984 (World Scientific, Singapore, 1986) p. 250; Geometry of nonlinear field theories (World Scientific, Singapore, 1986) p. 157.
[6] É. Cartan, *Leçons sur la géométrie des espaces de Riemann* (Gauthier-Villars, Paris, 1963) 2nd ed., pp. 177-184.
[7] E. Witten, *Nucl. Phys.* **B266** (1986) 245;
M.T. Grisaru, P. Howe, L. Mezincescu, B. Nilsson, and P.K. Townsend, *Phys. Lett.* **162B** (1985) 116;
E. Bergshoeff, E. Sezgin, and P.K. Townsend, *Phys. Lett.* **169B** (1986) 191.
[8] J.A. Shapiro and C.C. Taylor, *Phys. Lett.* **181B** (1986) 67, **186B** (1987) 69; *Phys. Reports* **191** (1990) 221.
[9] M.T. Grisaru, H. Nishino, and D. Zanon, *Phys. Lett.* **206B** (1988) 625, *Nucl. Phys.* **B314** (1989) 363.
[10] S.J. Gates, Jr., *Phys. Rev.* **D17** (1978) 3188;
M. Brown and S.J. Gates, Jr., *Annals Phys.* **122** (1979) 443;
S.J. Gates, Jr., M.T. Grisaru, M. Roček, and W. Siegel, Superspace, *or* One thousand and one lessons in supersymmetry (Benjamin/Cummings, Reading, 1983), p. 35.

[11] W. Siegel, *Nucl. Phys.* **B142** (1978) 301;
S.J. Gates, Jr. et al., *loc. cit.*, p. 280.
[12] N. Berkovits, *Phys. Lett.* **304B** (1993) 249.
[13] M. Roček, private communication.
[14] V.I. Ogievetskii and E. Sokatchev, *Phys. Lett.* **79B** (1978) 222; *Sov. J. Nucl. Phys.* **31** (1980) 140, 424, **32** (1980) 447, 589.
[15] E. Bergshoeff, P.S. Howe, C.N. Pope, E. Sezgin, and E. Sokatchev, *Nucl. Phys.* **B354** (1991) 113.
[16] A. Giveon and M. Porrati, *Nucl. Phys.* **B355** (1991) 422.